# Electrical resistivity across the tricriticality in itinerant ferromagnet

P. Opletal[1], J. Prokleška[1], J. Valenta[1] and V. Sechovský[1]

[1]*Faculty of Mathematics and Physics, Department of Condensed Matter Physics, Charles University, Ke Karlovu 5, 121 16 Praha 2, Czech Republic*

**Abstract**

We investigate the discontinuous ferromagnetic phase diagram near tricritical point in $UCo_{1-x}Ru_xAl$ compounds by electrical resistivity measurements. Separation of phases in $UCo_{0.995}Ru_{0.005}Al$ at ambient pressure and in $UCo_{0.990}Ru_{0.010}Al$ at pressure of 0.2 GPa and disappearance of ferromagnetism at 0.4 GPa is confirmed. The exponent of temperature dependence of electrical resistivity implies change from Fermi liquid behavior to non-Fermi liquid at 0.2 GPa and reaches minimum at 0.4 GPa. Our results are compared to results obtained on the pure UCoAl and explanation for different exponents is given.

**Introduction**

Quantum phase transitions in metallic ferromagnet and especially weak ones have become interesting research area in recent years. With improvements in the sample preparation and development in theory new phenomena appeared. Experiments and theory were summarized in recent review by Brando et al.[1] One of the four discussed scenarios exhibits a discontinuous transition with tricritical point (TCP) and is assumed to be present in the pure systems. At TCP the ferromagnetic phase transition changes from second order to first order type and quickly disappears as a function of the tuning parameter. At TCP the first order metamagnetic transition from paramagnetic to ferromagnetic phase emerges in finite magnetic fields, usually denoted as metamagnetic "wings" in the phase diagram. The line connecting TCP and the endpoint at T=0 (denoted as quantum wing critical point) follow the crossover between first order transition at lower temperatures and second order transition at higher temperatures.

One of the compounds exhibiting part of the above mentioned type of the phase diagram is UCoAl.[2] It is an archetypal 5f-electron itinerant metamagnet, crystallizing in hexagonal ZrNiAl structure. TCP was estimated to be in negative hydrostatic pressures[3] for pure stoichiometric compound. Ferromagnetism can be induced by doping (e.g. Ru on Co site[4]) or small off-stoichiometry.[5] Recently,[6] a separation of phases in $UCo_{0.995}Ru_{0.005}Al$ at ambient pressure and together with the change induced by hydrostatic pressure from ferromagnetic to paramagnetic phase through TCP in $UCo_{0.990}Ru_{0.010}Al$ have been observed. In this work we extend the experiments by means of acquisition and analysis of electrical transport data and theirs comparison to published magnetic and transport data for UCoAl, UCo$_{0.995}$Ru$_{0.005}$Al and UCo$_{0.990}$Ru$_{0.010}$Al.

**Experimental**

The preparation and characterization of single crystals of UCo$_{0.990}$Ru$_{0.010}$Al and UCo$_{0.995}$Ru$_{0.005}$Al is described in detail in Ref. 6. Thin rectangular-shape samples have been prepared for four probe electrical resistivity measurement. Measurements were done in a PPMS 9 (Quantum Design) and dilution refrigerator (Leiden Cryogenics). A double layered CuBe/NiCrAl pressure cell[7] with Daphne 7474 as the pressure medium was used. Pressure was determined by manganin manometer at room temperature and corrected for the temperature evolution of pressure. The pressures mentioned in the paper are stated for low temperatures. In order to obtain only the diagonal term of the resistance tensor, the electrical resistivity was measured in a positive and negative magnetic fields and only the symmetric part was considered (to exclude the mixing with the Hall effect).

**Results and discussion**

In the previous study[6] the coexistence of ferromagnetic and metamagnetic phase was found in UCo$_{0.995}$Ru$_{0.005}$Al. The ground state of UCo$_{0.995}$Ru$_{0.005}$Al is ferromagnetic, Curie temperature $T_C$~ 4 K. Above this temperature the mentioned coexistence appears up to ~ 9 K.[6] The electrical resistivity and its temperature derivative (Fig. 1) at zero magnetic field shows no clear indication of transition around 4 K. This absence could be explained by weak nature of the first order transition and significant presence of spin fluctuations, both above and below Curie temperature. The maximum of the derivative of electric resistivity coincides with maximum in thermal expansion,[6] connected to the critical endpoint. Fitting measured data by power law dependence $\rho = \rho_0 + AT^n$ in ferromagnetic regime (up to 4 K) gives $n = 2$, typical for Fermi liquid behavior. Plotting resistivity against $T^2$ (Fig. 1), we see the faster increase of the electrical resistivity above $T_C$ (as compared to $T^2$ in ferromagnetic regime). This faster increase terminates around 8 K as seen in the resistivity derivative in Fig. 1, too.

The magnetic field dependence of the electrical resistivity of UCo$_{0.995}$Ru$_{0.005}$Al at low temperatures is similar to that for pure UCoAl (Fig. 2). In UCoAl the negative step growing with the increasing temperatures is observed in transversal setup similar to our sample. This anomaly is connected to the metamagnetic transition, its origin was discussed in Ref. 8. Contrary to the observations in the magnetization measurements,[6] the nucleation of the metamagnetic transition is not significantly hindered by the presence of saturated magnetization and continuous increase (both in position and step size) is observed from the lowest field. It should be noted that in our case another contribution to the temperature dependence of the step size has to be considered – in the mixed region the increasing amount of the metamagnetic phase will increase the step size as well. The positive magnetoresistance at higher temperatures is of magnetic origin, in particular it is attributed to growth of magnetic moments below critical field.

The negative incline in high magnetic fields comes from suppression of spin fluctuations similar to the case of pure UCoAl.[8]

UCo$_{0.995}$Ru$_{0.005}$Al provide us insight only in one cut of the phase diagram. To examine the rest of the phase diagram we measured UCo$_{0.990}$Ru$_{0.010}$Al, which is well-defined ferromagnet with $T_C$ = 16 K, and follow the effect of suppression of ferromagnetism[6] on transport properties as a function of hydrostatic pressure. From Fig. 3 we see the knee in the temperature dependence of electrical resistivity at ambient pressure due to the phase transition. This knee is smeared out with increasing pressure and the temperature dependence remains featureless, similarly to UCo$_{0.995}$Ru$_{0.005}$Al at ambient pressure. To investigate the pressure evolution we use again the formula $\rho = \rho_0 + AT^n$ (see Fig. 4). In ambient pressure the best agreement is observed for n = 2, characteristic for the Fermi liquid behavior as expected for ferromagnet below the ordering temperature. Interestingly, in low hydrostatic pressures the agreement with n = 2 is observed only at low temperatures, even though sample is ferromagnet up to significantly higher temperatures. The change in behavior can be explained by appearance of other scattering term (the TCP is at ~0.2 GPa[6]). The effect of TCP is seen in the change of coefficient $A$ (incline of line), too. At 0.12 GPa $A$ starts to change above 6 K, while at 0.17 GPa closeness to TCP affects the whole measured temperature interval. Data in this pressure for lower temperature are desirable. At 0.22 GPa and higher pressures we can find a better agreement for $n=5/3$ and especially $n=3/2$ in highest pressures signaling non-Fermi liquid (nFl) behavior.

Comparing our results to those for pure UCoAl we observe a similar value of the A-coefficient, but different exponent $n$. While the electrical resistivity of pure UCoAl behaves as ~ $T^{5/3}$,[9] both our samples behaves as $T^{3/2}$. Recently Kirkpatrick and Belitz[10] explained $T^{3/2}$ behavior by existence of static droplets of ordered phase in magnetically disordered phase. Applying the proposed scenario lead us to conclusion, that although the UCoAl at ambient pressure and UCo$_{0.990}$Ru$_{0.010}$Al under applied hydrostatic pressure have the same magnetic ground state,[6] the microscopic nature of this state is different, and reflects the paths necessary to reach such a state. In the case of UCoAl, the compound has a well ordered crystal structure and its electronic configuration places it at the verge of magnetism resulting in an archetypal metamagnet. On the other hand, the substituted sample, even forced to the same state reflects its history with the irregularities in the lattice acting as a condensation nuclei leading to the survival of the nFl behavior (reminiscence of TCP) across the extended range in the parameter space. With respect to this result a detailed study comparing the pure UCoAl and doped samples at even higher pressures is desirable.

Temperature dependence of magnetoresistivity for both samples and in various hydrostatic pressures is depicted in Fig. 5. For UCo$_{0.990}$Ru$_{0.010}$Al we observe a minimum in all cases. At ambient pressure the minimum is found at $T_C$ and shows a clear presence of long range order, being not affected by the vicinity of the TCP. From 0.12 GPa up to 0.38 GPa the temperature of the minimum is constant and the minimum becomes more shallow with increasing pressure. This is explained by the presence of the mixed-phase region in the vicinity of the TCP. At 0.45 GPa minimum starts to shift to lower temperatures and at 0.72 GPa two minima are indicated. A similar high temperature minimum has been observed for pure UCoAl and related to the maximum in magnetic susceptibility χ(T) which is bounded to spin fluctuations as in other metamagnets.[8]

Vanishing of the minimum with increasing pressure and its constant temperature suggests its origin in the ferromagnetic portion of the mixed phase. The mixed phase is suppressed above 0.4 GPa and at highest pressures the minimum can be explained by the same mechanisms as in pure UCoAl - critical magnetic field at 0.72 GPa is close to 3 T (field at which resistivity was measured) and the low temperature minimum is due to crossover from paramagnetic to ferromagnetic phase.

Interestingly $UCo_{0.995}Ru_{0.005}Al$ exhibits the minimum at the same temperature as $UCo_{0.990}Ru_{0.010}Al$ (Fig. 5) in pressures between 0.12 GPa and 0.38 GPa. This is in agreement with $UCo_{0.995}Ru_{0.005}Al$ being in the mixed phase part of phase diagram.

For completeness it should be noted, that the same behavior is observed in the field dependencies of resistivity measured at fixed temperatures at different pressures – in purely ferromagnetic state (low temperatures, low pressures) resistivity monotonously decreases with increasing magnetic field. In medium pressures the mixed phase magnetoresistivity has the same form as in $UCo_{0.995}Ru_{0.005}Al$ (see Fig 2 and discussion earlier). At high pressures in the paramagnetic phase the increase in low magnetic fields is observed, followed by maximum and negative decline afterward. The increase is attributed to the magnetic scattering coming from increasing of the magnetic moment – with increasing pressure the shape of the peak starts to change to plateau, similarly to UCoAl.[11] Kimura et al.[2] suggested the existence of a supercritical region where additional scattering appears and explains this plateau. It is worth to mention that no similar anomaly appears in longitudinal setup. In higher field we again observe negative magnetoresistivity due to the suppression of the spin fluctuations.

**Conclusions**

The obtained data confirm the phase separation in $UCo_{0.995}Ru_{0.005}Al$ and loss of magnetism with hydrostatic pressure of $UCo_{0.990}Ru_{0.010}Al$ as proposed in Ref. 6. In the latter case, the electrical resistivity exponent *n* changes from 2 (Fermi liquid) in pure ferromagnetic state to 3/2 in the paramagnetic ground state with increasing pressure. Although it is well established[6] that the doped system under high pressure shows the same magnetic ground state as pure UCoAl at ambient pressure, the subtle differences in the microscopic nature are reflected by different power law, as the pure UCoAl follows n = 5/3.[2] In the transport properties the presence of the TCP is reflected in the position and strength of the magnetoresistance minimum.

**Acknowledgments**

This work is a part of the research program GACR 16-06422S, which is financed by the Czech Science Foundation and GAUK 363015, which is financed by Grant Agency of Charles University.

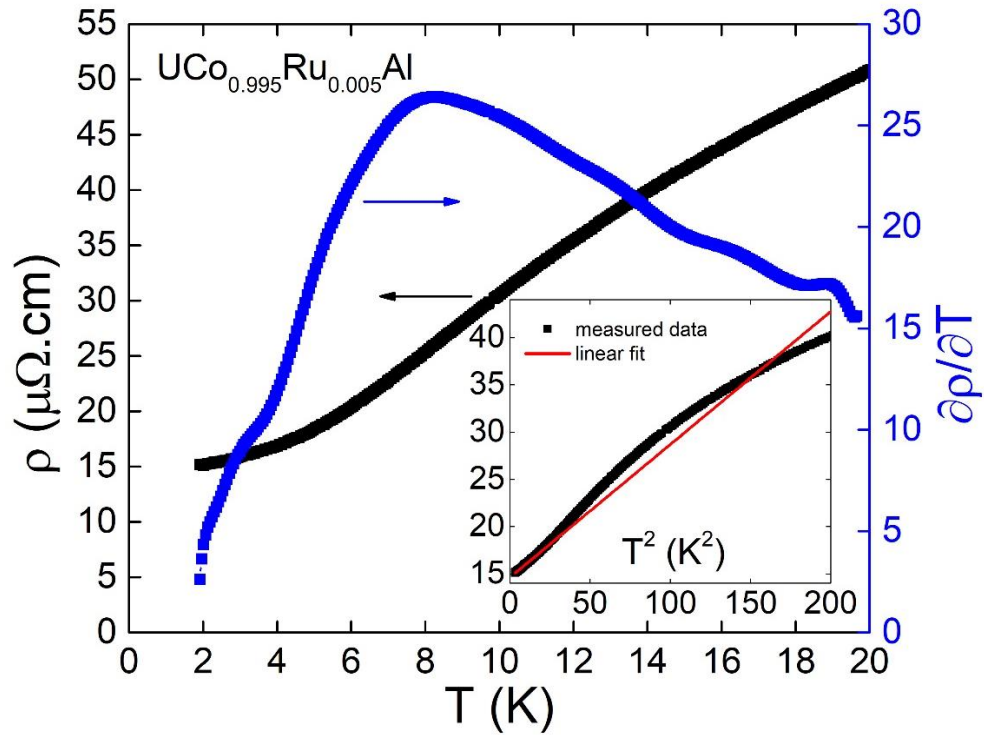

Figure 1: Temperature dependence of the electrical resistivity (black) and its derivative (blue) of UCo$_{0.995}$Ru$_{0.005}$Al with electrical current in basal plane.

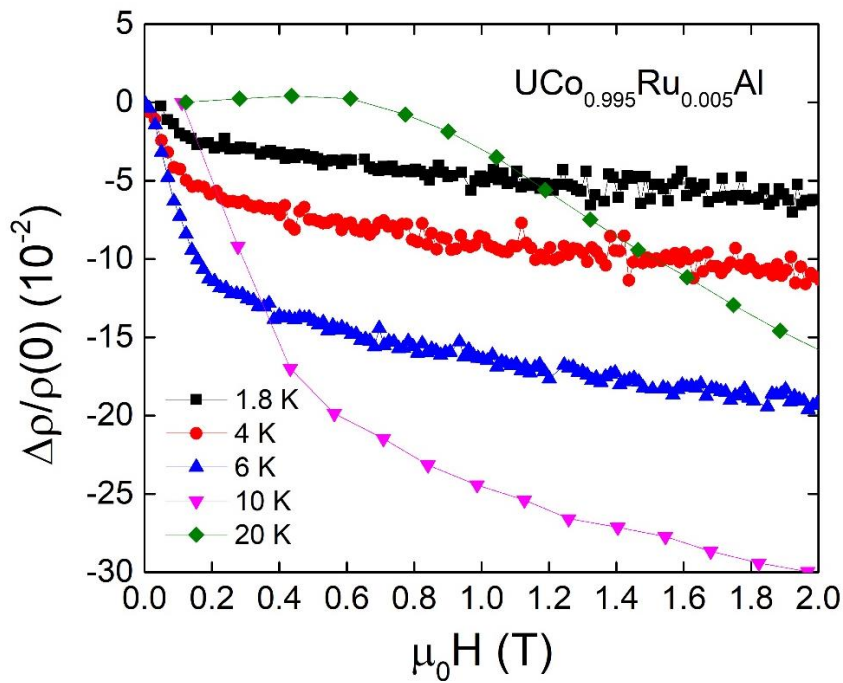

Figure 2: Magnetoresistivity of UCo$_{0.995}$Ru$_{0.005}$Al at different temperatures in transversal setup for field applied along the c-axis

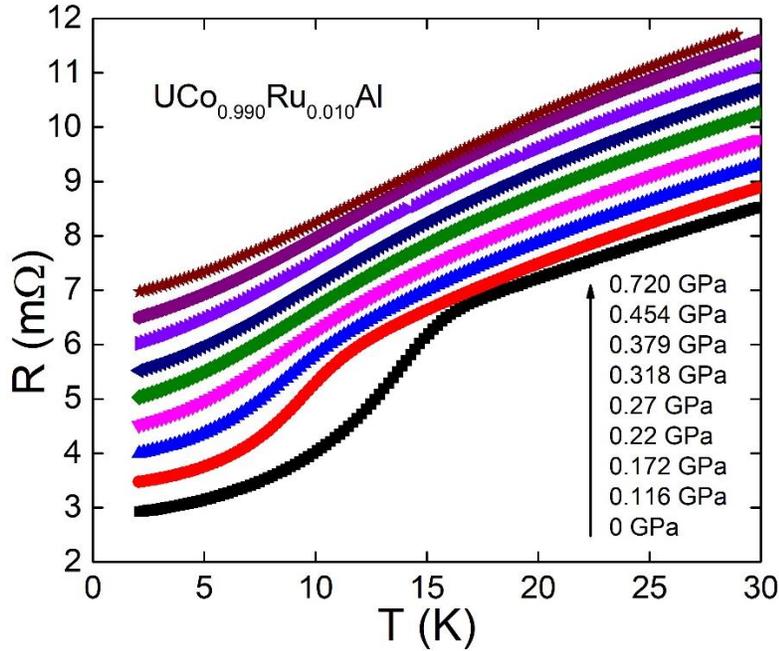

Figure *3*: Temperature dependence of electrical resistivity of UCo$_{0.990}$Ru$_{0.010}$Al at different pressures in transversal setup. Data for each pressure are shifted for better clarity.

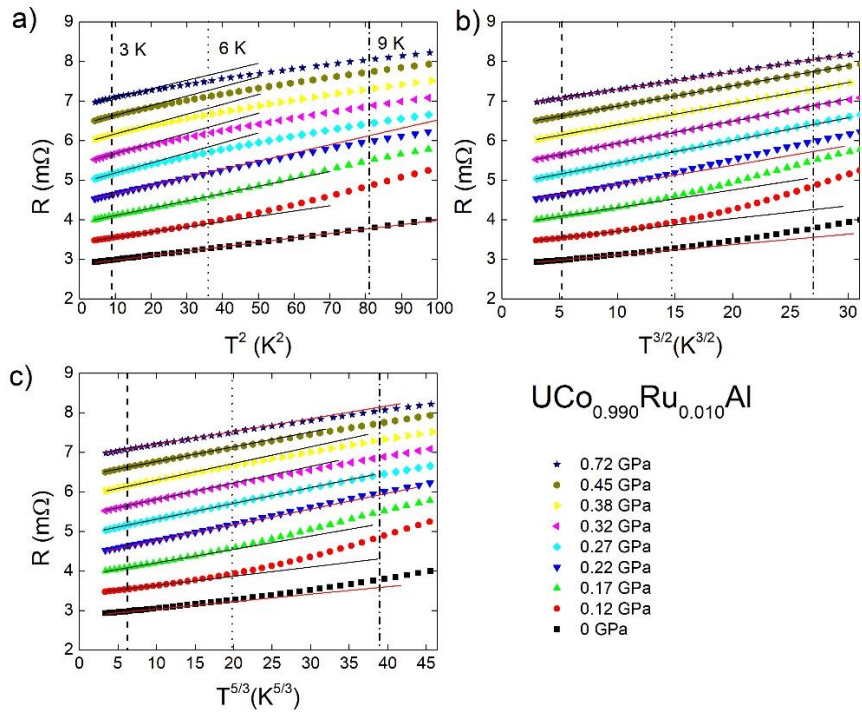

Figure *4*: Electrical resistivity plotted against $T^n$ for UCo$_{0.990}$Ru$_{0.010}$Al in different pressures. Data are shifted for better clarity. Lines are linear fits in low temperatures are chosen for better visibility. The vertical lines denote the same temperatures in all three panels.

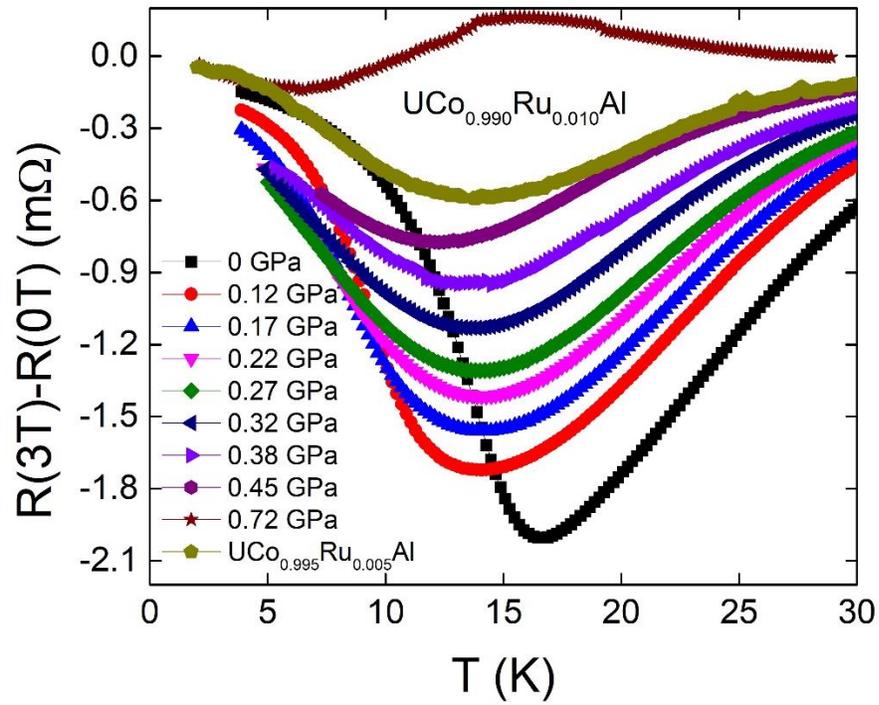

Figure 5: Temperature dependence of magnetoresistivity of UCo$_{0.990}$Ru$_{0.010}$Al in different pressures and UCo$_{0.995}$Ru$_{0.005}$Al at ambient pressure in transversal setup for field applied along the c-axis.